\documentclass[preprint]{aastex}

\slugcomment{Accepted 28 June 2010 to appear in {\it ApJ Supplements}}

\shorttitle{HST Astrometry}
\shortauthors{Benecchi et al.}

\begin{document}


\title{HST Astrometry of Transneptunian Objects}


\author{S. D. Benecchi\altaffilmark{1} and K. S. Noll\altaffilmark{2}}
\affil{Planetary Science Institute, 1700 East Fort Lowell, Suite 106, Tucson, AZ 85719}


\altaffiltext{1}{Planetary Science Institute, 1700 East Fort Lowell, Suite 106, Tucson, AZ 85719; susank$@$psi.edu}
\altaffiltext{2}{Space Telescope Science Institute, 3700 San Martin Dr., Baltimore MD 21218}


\begin{abstract}
We present 1428 individual astrometric measurements of 256 Transneptunian objects made with HST. The observations were collected over three years with two instruments, the Wide Field Planetary Camera 2 and the Advanced Camera for Surveys High Resolution Camera, as part of four HST programs. We briefly describe the data and our analysis procedures. The submission of these measurements to the Minor Planet Center increased the individual arc length of objects by 1.83 days to 8.11 years. Of the 256 total objects, 62 (24.2${\%}$) had arc length increases $\ge$ 3 years. The arc length for 60 objects (23.4${\%}$) was increased by a factor of two or greater.
\end{abstract}


\keywords{Kuiper belt: general, Astrometry}



\section{Introduction}

Where objects reside in the Kuiper Belt provides constraints for solar system formation models (Gomes et al. 2005; Morbidelli et al. 2005; Tsiganis et al. 2005; Levison et al. 2007). As of June 2010 there are 1418 objects cataloged by the Minor Planet Center (MPC). These objects reside in a variety of orbits characterized by their orbital elements and their interaction, or non-interaction, with Neptune. Two classification schemes have been suggested: (1) the Deep Ecliptic Survey (DES; Chiang et al. 2003; Elliot et al. 2005), and (2) the Gladman classification (Gladman et al. 2008) with the greatest distinction being the boundary between the Classical and Scattered objects. For the purposes of this paper we follow the DES scheme (Elliot et al. 2005) with more recent enhancements to include the ÒscatteringÓ concept from Gladman et al. (2008). Objects with orbits in mean motion resonances with Neptune are classified as Resonant. Objects with low inclination and low eccentricity orbits that do not interact with Neptune are called Classical KBOs. Objects with highly eccentric  ($\ge$0.2) and highly inclined orbits are known as Scattered disk objects. The Centaurs, objects on short-lived giant-planet-crossing orbits can be consider to be a subset of the Scattered disk. The objects we report on in this paper are nearly equally split between the three main populations of Transneptunian Objects (TNOs) including objects sampling 20 unique mean motion resonance locations.

Most of the known TNOs have been discovered by dedicated surveys. In such surveys two or more images are taken 2-24 hours apart and TNOs are identified as slow moving objects (3-5 arcseconds/hour) in the field (mostly due to the reflex motion of the Earth). Follow-up observations months and years later refine the orbit of an object so that it can be routinely found again and dynamically classified. 

TNO orbital periods are roughly 250 years or more (Pluto's period is 248.09 years), increasing with heliocentric distance.  It takes $\sim$3 years to refine an orbit for dynamical classification which is most sensitive to the determination of the semi-major axis and eccentricity (Wasserman et al. 2006). For example, for the objects where our astrometry measurements extended the arc by the greatest fractional length (these objects included a variety of dynamical classes) we explored their orbit space with the Bernstein formalism (Bernstein \& Khushalani, 2000) and theoretical measurements at a variety of time intervals (approximately two measurements near opposition separated by a month during year one, the same type of observation in year 2 and a single observation near opposition in year 3). We estimate that with observations spanning the first lunation after discovery of a TNO, the semi-major axis is determined to an accuracy of $\sim$20\% and the eccentricity to $\sim$80\%. The uncertainty drops to $\sim$2\% in semi-major axis and $\sim$10\% in eccentricity after recovery in the second year and to less than 1\% in both semi-major axis and eccentricity following observations in year 3. The observational uncertainty of an object's position after three years of observations is approximately 0.5 arcseconds. The positional uncertainty increases by approximately 0.5 arcseconds each year. For additional discussion of orbit determination and uncertainties see Jones et al. 2010.

Observations with HST require the positional uncertainties of a TNO to be within the relatively modest field of view (FOV) of the camera (on the order of 10s of arcseconds). Most of the objects we observed already had positional uncertainties significantly better than this limit. The primary goal of the observing programs was to look for binaries. The main requirement for observation was that the object have a predicted visual magnitude brighter than $\sim24$ and an expected positional accuracy within the field of view of the HST instrument ($\sim$25 arcseconds for ACS/HRC and $\sim$35 arcseconds for WFPC2). The objects we report here verify earlier astrometry and extend the duration of comparably accurate astrometry, in almost a quarter of the cases by three years or more. There is continual need for improved astrometry measurements for both photometric and spectroscopic follow-up in small field imagers as well as in the future for occultation studies. Clearly our measurements alone will not refine the orbits enough for targeted occultation predictions (needing positions good to 10s of milliarcseconds), however it is important to note that at some point the noise floor for a TNO's orbit is dominated by errors in the astrometric reference catalog, not by the precision of the orbital elements themselves.  It is suggested that this level is reached for a typical TNO when the orbit arc reaches approximately eight years (M. Buie, private communication, 2010). TNOs targeted for occultation studies will likely require focused astrometric campaigns where both the TNO orbit and star positions in the field of interest can be simultaneously improved.  For 38 objects our observations extended the arc lengths to 8 years or more.

\section{Observations}


In this work we present astrometry measurements from four HST programs that observed TNOs for binary identification and photometric/binary orbit studies. The details of the proposals and specific results can be found in Stephens et al. 2006, Noll et al. 2006, 2008, Grundy et al. 2007, 2008, 2009, and Benecchi et al. 2009.  The data were obtained from HST programs 10514, 10800, 11113 and 11178 and included 256 unique objects which were observed 3-4 times in a single HST orbit with the telescope tracking at the rate of the TNO. A few objects were observed in multiple HST orbits. The programs which searched for binaries selected objects whose visible magnitudes were brighter than 24 and whose expected positional uncertainties were within the field of view of the camera being used at the time. The timing of these observations was not specific; data were collected when it was convenient to do so with the telescope. The programs aimed at photometric and binary studies were specifically targeted in both time and object.   Observations were obtained using the Advanced Camera for Surveys High Resolution Camera (HRC) and the Wide Field Planetary Camera 2 (WFPC2). All of the observations were dithered to avoid fixed pattern noise, properly sample the point-spread function (PSF), and provide for better statistics.  Exposures were 260-300 seconds in duration and were obtained using a broad filter (Table 1) so as to obtain the best throughput possible for each object. 




\section{Data Reduction and Analysis}

\subsection{HST Pipeline Processing}

We processed the data through the standard HST pipeline (Baggett et al. 2002; Pavlovsky et al. 2006). The basic image reduction flags static bad pixels, performs A/D conversion\footnote{A/D conversion takes the observed charge in each pixel in the CCD and converts it to a number with the appropriate gain setting. The gain for ACS/HRC is 2.216. }, subtracts the bias, and dark images, and corrects for flat fielding. It also updates the header with the appropriate photometry keywords.  The flat-field-calibrated images, subscripted by {\it flt} for ACS and {\it c0f} for WFPC2, were the ones we analyzed.

\subsection{Astrometry and Photometry}

The position for each TNO was determined using the Point Spread Function (PSF) fitting method described in Benecchi et al. (2009). These positions were then used in combination with the guide star positions from the HST image headers and the appropriate geometric distortion corrections to obtain the Right Ascension ($\alpha$) and Declination ($\delta$) of the TNO in the field. In the following paragraphs we elaborate on the details of our analysis methods.

PSF fitting provides the most accurate position of the center of light of the TNO in the HST images. Our analysis began by identifying the TNO on the image (the only non-smeared object in the field since the telescope was tracking at the rate of the TNO) and obtaining a crude centroid (chosen by eye, good to about 0.5 pixel), flux (summed within a 1.5 pixel radius around the chosen center) and sky background estimation. If the TNO was a binary, we reported the position as if for a single object. For close binaries the position is effectively the center of light, for well resolved binaries the position is that of the primary.  Next, a sub-pixel sampled model PSF was generated using Tiny Tim [generated for a solar spectral distribution using models 51 (HD 150205, a G5V star) and 57 (HD 154712, a K4V star) from the Bruzual-Persson-Gunn-Stryker Spectra Library; Krist \& Hook 2004] with the crude position, flux and sky background characteristics.  The model was subtracted from the data and a $\chi^{2}$ was calculated to provide a reference point. Next we modified the PSF to account for changes in focus (thermally-induced ÒbreathingÓ) and for small motions of the spacecraft known as ÒjitterÓ that occur on orbital timescales. We iterated the determination of the focus and jitter values with an automated fitting routine, {\it amoeba} [Press et al. 1992; this routine performs multidimensional minimization of a function containing our object variables (x, y, background and flux) using the downhill simplex method], until the $\chi^{2}$ converged. Typically, 4-5 iterations were required to reach a final PSF model. 

From our fits, we obtained the nominal x and y position of the TNO on the HST image ({\it flt} or {\it c0f}). The next piece of information required for astrometry is the position of the reference stars. We used the guide star position values (referenced in the world coordinate system (WCS), Pavlovsky, C. et al. 2006, section 6.2)\footnote{The keywords for the reference position are found under the heading "World Coordinate System and Related Parameters" in the .fits headers. The values from the keywords {\it CRPIX1}, {\it CRPIX2}, {\it CRVAL1}, {\it CRVAL2}, {\it CD1\_1}, {\it CD1\_2}, {\it CD2\_1} and {\it CD2\_2} are used in the PYRAF script.}  within the header of the image, combined with the geometric distortion table for the HRC (found in the {\it idc} files) and the PYRAF {\it xytosky} routine to obtain the ($\alpha, \delta$) of the TNO in the field. The process was similar for the WFPC2 image with the exception of the application of the geometric distortion corrections, of which there are none\footnote{When the analysis for this paper was done the WFPC2 images needed to be converted from GEIS to FITS image format prior to extracting the astrometry. The conversion was accomplished with the IRAF routine strfits. Since the time of this work, the HST archive was updated to provide all images in standard FITS format.}.  The positions of the guide stars are accurate to $\sim1$ arcsecond (Pavlovsky, C. et al. 2006, section 6.2.2). Most of our TNOs were observed near the center of the image where the astrometry is the most accurate, however, the uncertainties for objects near the edges of the image are slightly larger.

For submission to the MPC, in addition to the ($\alpha, \delta$) of the TNO, the position of HST in its x-,y-,z-axis at the time of the observation is required. This information is contained within the archive table files with extensions {\it spt} for HRC and {\it shf} for WFPC2. The reference keywords are {\it POSTNSTX}, {\it POSTNSTY} and {\it POSTNSTZ} for HST position X-axis, Y-axis and Z-axis, respectively. The units of the values are in km and they are given to an accuracy of 8 decimal places. Added in quadrature these values give the instantaneous distance of HST from the Earth's center ($\sim$6942 km). 

As a final step, photometry was performed on the images to provide a magnitude for each object at the time of observation. We followed the same process described in Benecchi et al. (2009). We summed the object counts on the scaled, noiseless PSF model within a 0.5 arcsecond aperture and converted to magnitude in the standard fashion.  Next, we corrected that value to an infinite aperture by adding an offset value from Table 5 in Sirianni et al. (2005). The photometric zero point was determined from the {\it photflam} and {\it photzpt} values in the image header which are given in the Space Telescope magnitude system (STmag). Finally, we used {\it synphot} to convert from the STmag system to the standard Vega magnitude system. 

After collecting astrometry and photometry information for each object and the position of HST at the time of each observation, files were constructed in the appropriate format for submission to the MPC.  The object names were converted to the ÒpackedÓ format required by the MPC and defined as: the first two digits of the year are a single character in column 1 (eg. I=18, J=19, K=20), columns 2-3 contain the last two digits of the year, column 4 contains the half-month letter and column 7 contains the second letter. The 0 to 3 number cycle count is placed in columns 5-6, using a letter in column 5 when the cycle count is larger than 2 digits (eg. A=10, B=11). For example, 04VS75 = K04V75S and 04XX190 = K04XJ0X. Similarly for a numbered object, if the number is less than 5 digits all the values are used. If the number is 6 digits the first two are converted to a letter (eg. A=10, B=11).  For example, 78799 = 78799 and 120181 = C0181 (reference MPC website/documentation). 

A sample of the text file for submission to the MPC as specified in documentation on the MPC website (http://www.cfa.harvard.edu/iau/info/NewObsFormat.eps) can be found in Figure 1, the file is space delimited. The astrometric results for all our submitted measurements can be found in Table 1.

\begin{figure}
\epsscale{.95}
\plotone{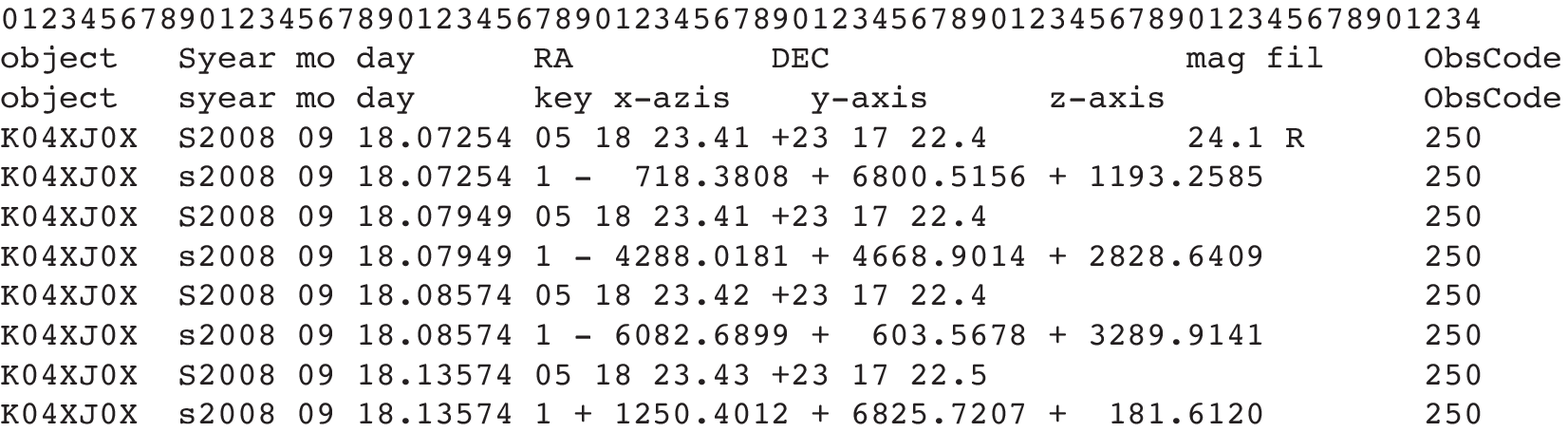}
\caption{{\bf Sample MPC file structure.} The MPC file is space delimited, the first line for an object gives the position of the TNO,  the second line gives the position of HST when the observation was taken.} \label{fig 1}
\end{figure}




\subsection{Orbit check and residuals}

After astrometry is compiled for all our observations we compare our measured position to the calculated position at the time of observation based on the orbit of the object in the MPC database. We call the difference between these two values, observed minus calculated, the ÒresidualÓ.  We observed both numbered, objects which the MPC has deemed to have reliable positions far into the future (typically observed at four or more oppositions), and unnumbered objects, objects which have been designated by the MPC with observations longer than 24 hours, but which don't have reliable long term orbits.  For numbered objects the residual is determined solely from the orbit in the MPC database.   For unnumbered objects, we calculate a new orbit using the Bernstein formalism (Bernstein \& Khushalani, 2000) and determine the residuals between our observations and the positions of the TNO for both the new and old orbit. In total, 1428 positions for 256 unique objects were submitted to the MPC. From the original sample there were 22 objects that we did not find in our images, likely due to bad magnitude estimates or poorer positional accuracy than expected. 

In most cases, since we were looking at objects with well-determined orbits, our observations do not significantly improve the orbits. However, in many cases our observations extend the arc on the orbit deminishing the chance the object will be lost in the future. Twenty-six percent of the objects we measured had positional improvements by 1.5-12 arcseconds; none were in the position of being lost (we define lost as having a positional uncertainty of $\gtrsim1800$ arcseconds which is an approximate limit for single field recovery with ground-based wide-field cameras on 4-m or larger telescopes) in the near future. In Table 1 we present the residuals for our measurements as provided by the MPC for a sample of objects with significantly improved arcs, the full table is provided electronically.  Column 1 is the object name in MPC format, columns 2, 3, and 4 are the time of observation in year, month and decimal day. Columns 5 and 6 are the $\alpha$ and $\delta$ position of the object, column 7 is the magnitude of the object in the respective filter, proposal ID and instrument (columns 8, 9 and 10). Columns 11 and 12 are the residuals (observed-reference) in $\alpha$ and $\delta$ space measured in arcseconds.  Residuals for numbered objects are referenced to the existing orbit and residuals for unnumbered objects are referenced to the new orbit. In columns 13 and 14 we list the increase of the arc length in years and the fractional arc length extension.

\begin{deluxetable}{rrrrrrrrrrrrrr}
\tabletypesize{\scriptsize}
\rotate
\tablecolumns{14}
\tablewidth{0pt}
\tablecaption{Astrometry (sample, full table electronic only)}
\tablehead{
\colhead{Object$^{a}$} & \colhead{Year}   & \colhead{Month}    & \colhead{Decimal Day} &
\colhead{RA (J2000.0)}    & \colhead{DEC (J2000.0)}   & \colhead{$M_{HST}$}    & 
\colhead{Filter}  & \colhead{PropID}  & \colhead{Instr} & 
\colhead{Resid ('')$^{b}$}  & \colhead{Resid ('')$^{b}$}  & \colhead{Prearc$^{c}$} & \colhead{Postarc$^{d}$}}
\startdata
  15820&    2008&      09& 27.57185 &$01:21:59.47$&$+21:16:52.7 $&   25.4&   F606W&   11113&   WFPC2& -2.95&  0.65&  13.908&   13.989\\
\nodata& \nodata& \nodata& 27.57879 &$01:21:59.43$&$+21:16:52.6 $&\nodata& \nodata& \nodata& \nodata& -2.71&  0.77& \nodata& \nodata \\
\nodata& \nodata& \nodata& 27.58504 &$01:21:59.37$&$+21:16:52.5 $&\nodata& \nodata& \nodata& \nodata& -2.79&  0.92& \nodata& \nodata \\
\nodata& \nodata& \nodata& 27.59199 &$01:21:59.34$&$+21:16:52.4 $&\nodata& \nodata& \nodata& \nodata& -2.41&  1.12& \nodata& \nodata \\
  15874&    2006&      07& 30.85538 &$03:32:08.05$&$+12:37:49.3 $&   20.3&   CLEAR&   10800&     ACS& -0.50&  0.06&   9.212&    9.805\\
\nodata& \nodata& \nodata& 30.85978 &$03:32:08.06$&$+12:37:49.3 $&\nodata& \nodata& \nodata& \nodata& -0.46&  0.06& \nodata& \nodata \\
\nodata& \nodata& \nodata& 30.86417 &$03:32:08.08$&$+12:37:49.3 $&\nodata& \nodata& \nodata& \nodata& -0.27&  0.10& \nodata& \nodata \\
\nodata& \nodata& \nodata& 30.86857 &$03:32:08.09$&$+12:37:49.2 $&\nodata& \nodata& \nodata& \nodata& -0.24&  0.07& \nodata& \nodata \\
  15875&    2008&      07& 22.18990 &$03:40:59.93$&$+25:15:38.3 $&   22.9&   F606W&   11113&   WFPC2& -0.62& -0.36&   7.033&   11.778\\
\nodata& \nodata& \nodata& 22.19615 &$03:40:59.97$&$+25:15:38.5 $&\nodata& \nodata& \nodata& \nodata& -0.43& -0.26& \nodata& \nodata \\
\nodata& \nodata& \nodata& 22.20240 &$03:41:00.01$&$+25:15:38.7 $&\nodata& \nodata& \nodata& \nodata& -0.30& -0.10& \nodata& \nodata \\
\nodata& \nodata& \nodata& 22.20865 &$03:41:00.05$&$+25:15:38.9 $&\nodata& \nodata& \nodata& \nodata& -0.27&  0.06& \nodata& \nodata \\
  19299&    2008&      10& 31.48430 &$02:28:30.57$&$+16:47:12.1 $&   23.9&   F606W&   11113&   WFPC2&  2.94&  0.98&   4.014&   12.124\\
\nodata& \nodata& \nodata& 31.49124 &$02:28:30.52$&$+16:47:11.9 $&\nodata& \nodata& \nodata& \nodata&  3.08&  1.02& \nodata& \nodata \\
\nodata& \nodata& \nodata& 31.55166 &$02:28:30.12$&$+16:47:10.0 $&\nodata& \nodata& \nodata& \nodata&  2.94&  0.89& \nodata& \nodata \\
\nodata& \nodata& \nodata& 31.55791 &$02:28:30.08$&$+16:47:09.9 $&\nodata& \nodata& \nodata& \nodata&  3.13&  1.01& \nodata& \nodata \\
  24952&    2008&      09& 27.50449 &$00:49:14.32$&$+14:42:41.8 $&   23.5&   F606W&   11113&   WFPC2& -3.34& -5.12&   3.022&   11.083\\
\nodata& \nodata& \nodata& 27.51074 &$00:49:14.28$&$+14:42:41.6 $&\nodata& \nodata& \nodata& \nodata& -3.25& -5.10& \nodata& \nodata \\
\nodata& \nodata& \nodata& 27.51768 &$00:49:14.24$&$+14:42:41.4 $&\nodata& \nodata& \nodata& \nodata& -3.07& -5.02& \nodata& \nodata \\
\nodata& \nodata& \nodata& 27.52463 &$00:49:14.20$&$+14:42:41.2 $&\nodata& \nodata& \nodata& \nodata& -2.94& -4.94& \nodata& \nodata \\
  24978&    2007&      05& 12.31003 &$16:00:35.04$&$-21:06:41.5 $&   23.1&   CLEAR&   10800&   WFPC2&  0.09& -1.03&   5.094&    9.037\\
\nodata& \nodata& \nodata& 12.31767 &$16:00:35.01$&$-21:06:41.4 $&\nodata& \nodata& \nodata& \nodata&  0.33& -1.09& \nodata& \nodata \\
\nodata& \nodata& \nodata& 12.37392 &$16:00:34.73$&$-21:06:40.6 $&\nodata& \nodata& \nodata& \nodata&  0.09& -0.96& \nodata& \nodata \\
\nodata& \nodata& \nodata& 12.38087 &$16:00:34.69$&$-21:06:40.5 $&\nodata& \nodata& \nodata& \nodata&  0.13& -1.00& \nodata& \nodata \\
  26181&    2008&      03&  9.95240 &$14:57:53.83$&$-09:49:32.4 $&   22.2&   F606W&   11113&   WFPC2& -0.38& -1.23&  24.164&   28.120\\
\nodata& \nodata& \nodata&  9.95865 &$14:57:53.82$&$-09:49:32.3 $&\nodata& \nodata& \nodata& \nodata& -0.23& -1.21& \nodata& \nodata \\
\nodata& \nodata& \nodata&  9.96490 &$14:57:53.81$&$-09:49:32.2 $&\nodata& \nodata& \nodata& \nodata& -0.09& -1.18& \nodata& \nodata \\
\nodata& \nodata& \nodata&  9.97115 &$14:57:53.79$&$-09:49:32.1 $&\nodata& \nodata& \nodata& \nodata& -0.14& -1.17& \nodata& \nodata \\
\enddata

\tablenotetext{a}{Object name in MPC packed format.}
\tablenotetext{b}{Provided by B. Marsden.}
\tablenotetext{c}{Orbit arc length in years previous to HST measurements.}
\tablenotetext{d}{Orbit arc length in years including HST measurements.}

\end{deluxetable}

 As an additional evaluation of our measurements, we plotted the residuals returned to us from the MPC to ensure that there are no systematics in our positions. As expected, we show in Figure 2 a random distribution for the residuals.  
 
\begin{figure}
\epsscale{.60}
\plotone{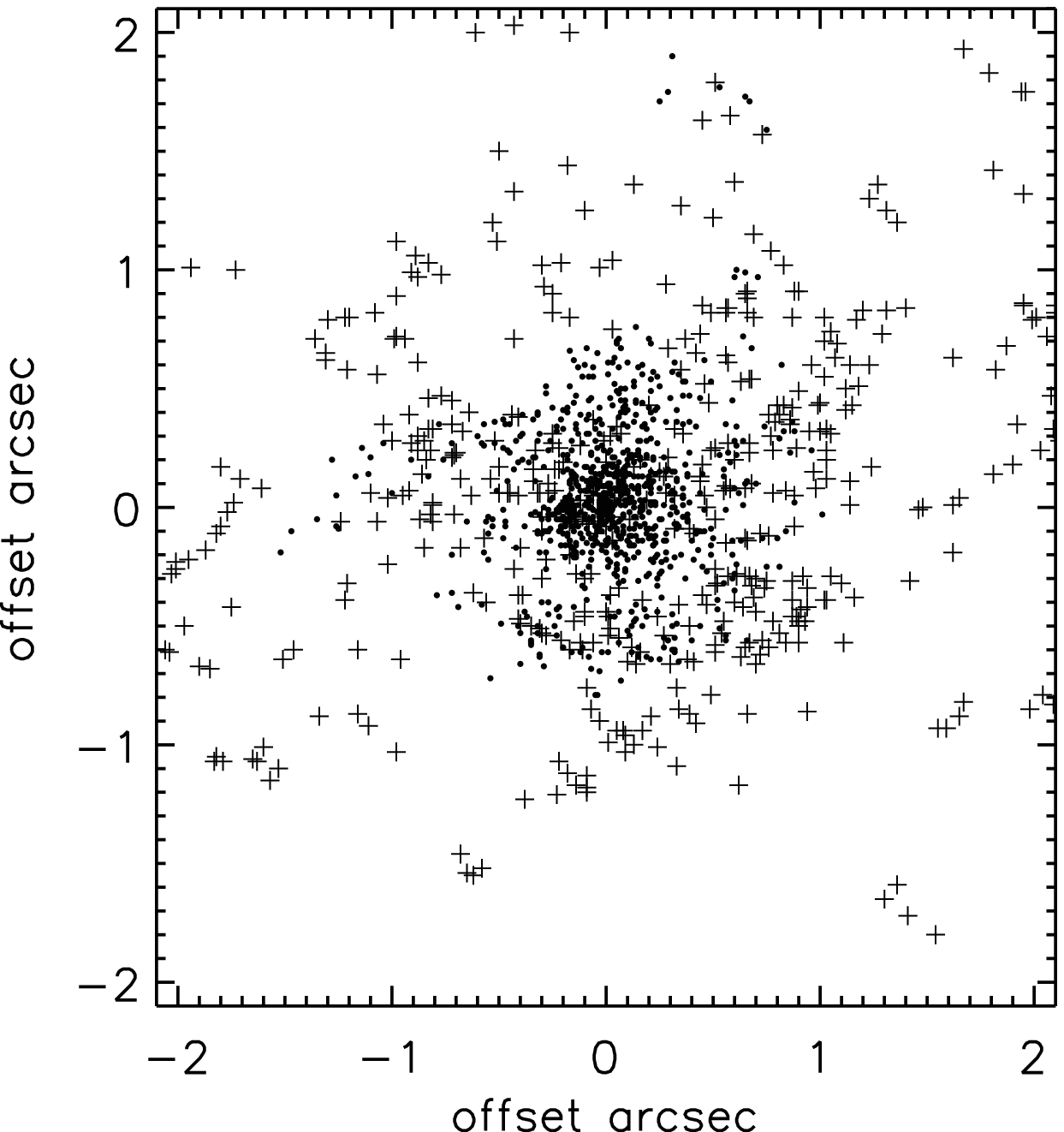}
\caption{{\bf Orbit residuals.} The difference between the observed position and expected position in the orbit of the TNO. Unnumbered objects are plotted as points and numbered objects are plus symbols. The numbered object residuals are based on orbits maintained by the MPC while the unnumbered object residuals are based on orbits calculated to include the new HST measurements. There do not appear to be any systematics among the residuals.} \label{fig2}
\end{figure}



\section{Discussion}


 In order to test the significance of the contribution of our astrometry, we investigated the orbits of the some of the objects in our survey that had the longest arc length increase (1997 $SZ_{10}$, 1998 $UR_{43}$, 1998 $WS_{31}$, 1998 $WY_{31}$, 19299 and 24952) and a similar number of objects which had the largest fractional arc length increase (2002 $CW_{224}$, 2000 $JF_{81}$, 2000 $YX_{1}$, 24952 and 19299). In both cases, we ran the Bernstein code with all the astrometric datapoints excluding the HST observations and once again with the HST astrometry included.  In all cases the objects had an arc length of at least 3 years previous to the addition of the HST datapoints. The addition of the HST datapoints do not change the nominal orbit solutions or the dynamical classifications. The dynamical classifications are calculated in as described in Elliot et al. (2005). The nominal orbit and the two orbits corresponding to 1-sigma extremes in semi-major axis and eccentricity space are integrated for $\sim10$ Myr with tests (in the order listed) for resonance membership, or for classification as a Centaur, Scattered or Classical object. All three orbits must give the same result for the object to be deemed classified.  In a few cases the semi-major axis changes on the order of 0.05 AU, but most of the time the changes are smaller than 0.01 AU. If an object were close to a resonance such an orbit adjustment might reclassify an object dynamically (resonant to non-resonant or vice versa), however we do not see any cases of that in our particular dataset.  It is also true that our observations decrease the uncertainities on the orbital elements (in particular those for semi-major axis, eccentricity and inclination) by a factor of $\sim$10 (e.g. from 0.01 AU to 0.001 AU on semi-major axis, from 0.0005 to 0.0001 on eccentricity and 0.001 to 0.0001 on inclination) for the objects with the largest fractional arc length increases.
 
Our results are in agreement with studies done by Elliot et al. (2005) and Wasserman et al. (2006) which show that observations over three years measured at two lunations on years one and two and at least one lunation in year three are sufficient to determine the orbit of a TNO to the accuracy required for finding the object in an typical CCD field for at least 10 years. For example, the object (49673) 1999 $RA_{215}$ was discovered in 1999 and has an arc length of 3.1 years. The current 1-year uncertainty on the orbit is 2.71 arcseconds.  To summarize the contribution of our observations we plot in Figure 3, a measure of the fraction of objects whose arc lengths were increased in year increments and the fractional arc length increases for all our objects. 
 
\begin{figure}
\epsscale{.90}
\plotone{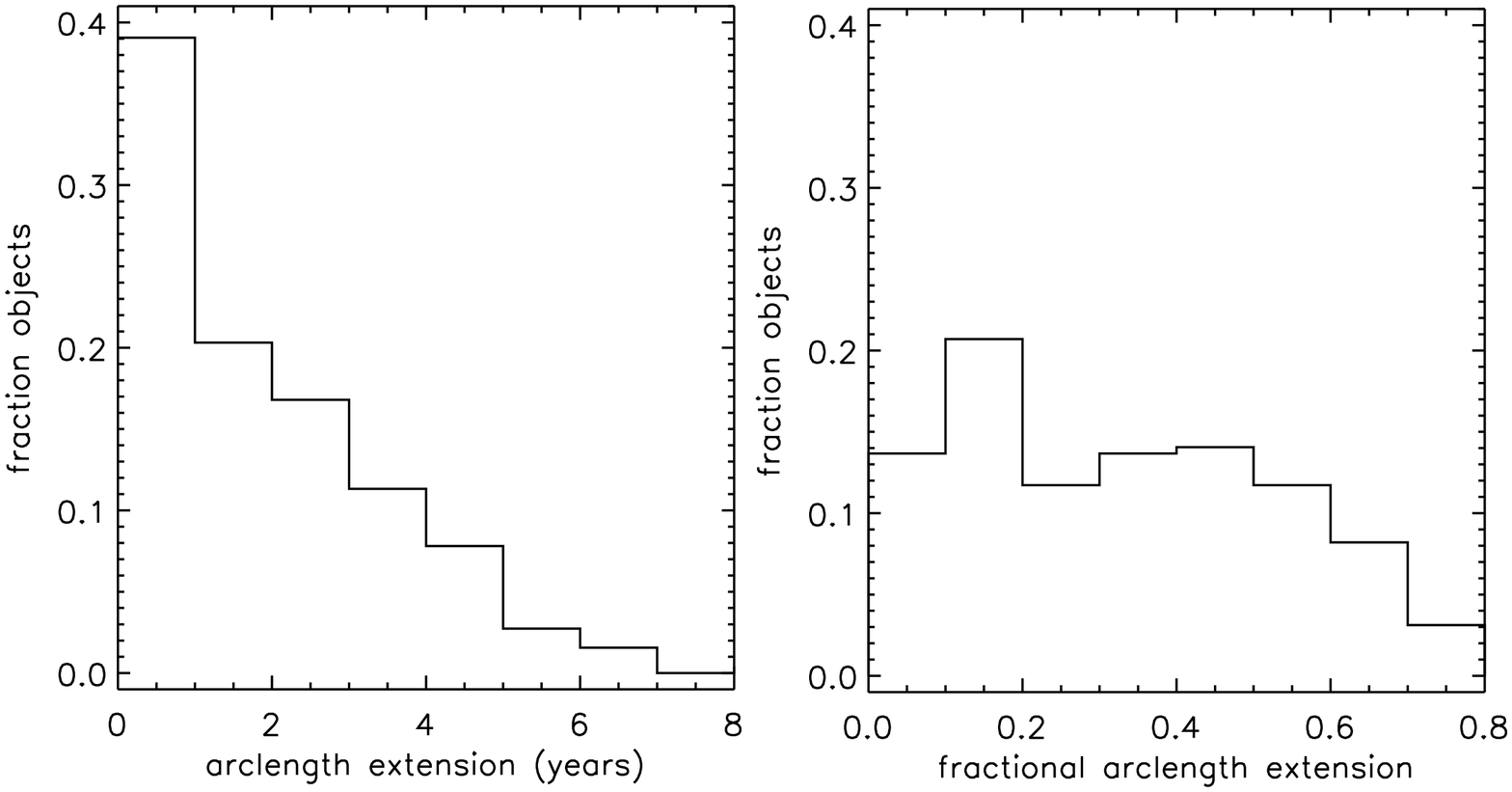}
\caption{{\bf Orbit Improvement.} (left) A plot of fraction of objects versus their arc length extension in years provided by the HST measurements.  Of the 256 total objects, 24.2$\%$ had their arc lengths extended by 3 years or more and 38 objects had their arc lengths extended beyond 8 years. (right) A plot of the fraction of objects versus their fractional arc length extension (i.e. arc length extension divided by total arc length). 23.4$\%$  of objects (60) had their fractional arc lengths improved by a factor of two or greater.}\label{fig3}
\end{figure}

 \section{Summary}

We have presented the steps required for extracting astrometry from HST images for submission to the MPC. The position of HST is a critical component of the measurement for moving objects in our solar system. Of the 256 TNOs measured, the individual arc length of objects were extended by 1.83 days to 8.11 years. 62 (24.2$\%$) objects had their arc length of observation increased by  $\ge$ 3 years.  An additional 38 objects now have orbital arc lengths greater than 8 years as a result of our measurements. 60 (23.4${\%}$) objects had fractional arc length improvements of a factor of two or greater.



\acknowledgments

This work is based on observations made with the NASA/ESA Hubble Space Telescope. These observations are associated with programs 10514, 10800, 11113 and 11178. Support for these programs was provided by NASA through a grant from the Space Telescope Science Institute, which is operated by the Association of Universities for Research in Astronomy, Inc., under NASA contract NAS 5-26555.



{\it Facilities:} \facility{HST (ACS, WFPC2)}.

\clearpage

{\bf REFERENCES: }
(1) Auri\`ere, M.  1982, \aap,  109, 301.
(2) Baggett, S., et al. 2002. HST WFPC2 Data Handbook: Version 4.0. B. Mobasher (Ed.), Baltimore, STScI.
(3) Benecchi, S. D. et al. 2009.  {\it Icarus}  200, 292-303.
(4) Bernstein, G., \& Khushalani, B. 2000. {\it AJ} 120, 3323-3333.
(5) Elliot, J. L. et al. 2005. {\it AJ} 129, 1117-1162.
(6) Gladman, B. et al. 2008. In: M. A. Barucci et al. (Eds.), {\it The Solar System Beyond Neptune}, Univ. of Arizona Press, Tucson, pp. 43-57.
(7) Grundy, W. M. et al. 2007. {\it Icarus} 191, 286. 
(8) Grundy, W. M. et al. 2008. {\it Icarus} 197, 260. 
(9) Grundy, W. M. et al. 2009. {\it Icarus} 200, 627.  
(10) Gomes, R. et al.  2005. {\it Nature} 435, 466-469.
(11) Jones, R. L. et al. 2010. {\it AJ} 139, 2249-2257.
(12) Levison, H. F. \& Morbidelli, A. 2007. {\it Icarus} 189, 196-212.
(13) Morbidelli, A. et al. 2005. {\it Nature} 435, 462-465.
(14) Noll, Keith S. et al, 2006. {\it Icarus} 184, 611.
(15) Noll, Keith S. et al. 2008. {\it Icarus} 194, 758.
(16) Pavlovsky, C. et al. 2006. HST Data Handbook for ACS, Version 5.0 (Baltimore: Space Telescope Science Institute).
(17) Stephens, D. C. et al.  2006. {\it AJ} 131, 1142.
(18) Tsiganis, K. et al. 2005. {\it Nature} 435, 459-461.
(19) Wasserman, L. H. et al. 2006. {\it BAAS} 38, 564.

\clearpage




\end{document}